\documentclass[aps,prl,reprint,twocolumn,amsmath,amssymb,floatfix,superscriptaddress,longbibliography]{revtex4-1} 

\usepackage{graphicx}  
\usepackage{dcolumn}   
\usepackage{bm}        
\usepackage{xcolor}
\usepackage[version=4]{mhchem}
\newcommand{\Eq}[1]{Eq.~(\ref{#1})}
\newcommand{\fig}[1]{Fig.~\ref{#1}}

\newcommand{\isec}{\rm{s}^{-1}}
\newcommand{\kd}{k_{\rm{des}} }
\newcommand{\ka}{k_{\rm{ads}} }

\begin{document}

\title{Anomalous protein kinetics on low-fouling surfaces}
\author{Mohammadhasan Hedayati}
\affiliation{Department of Chemical and Biological Engineering, 
Colorado State University, Fort Collins, CO 80523, USA}
\author{Matt J. Kipper}
\affiliation{Department of Chemical and Biological Engineering,
  Colorado State University, Fort Collins, CO 80523, USA}
\affiliation{School of Biomedical Engineering, Colorado State University,
  Fort Collins, CO 80523, USA}
\affiliation{School of Advanced Materials Discovery, 
Colorado State University, Fort Collins, CO 80523, USA}
\author{Diego Krapf}
\affiliation{School of Biomedical Engineering, Colorado State University,
  Fort Collins, CO 80523, USA}
\affiliation{School of Advanced Materials Discovery, 
Colorado State University, Fort Collins, CO 80523, USA}
\affiliation{Department of Electrical and Computer Engineering,
  Colorado State University, Fort Collins, CO 80523, USA}
\email[Corresponding author: ]{diego.krapf@colostate.edu}
\begin{abstract}
In this work, protein-surface interactions were probed in terms of adsorption and desorption of a model protein, bovine serum albumin, on a low-fouling surface with single-molecule localization microscopy. Single-molecule experiments enable precise determination of both adsorption and desorption rates. Strikingly the experimental data show anomalous desorption kinetics, evident as a surface dwell time that exhibits a power-law distribution, i.e. a heavy-tailed rather than the expected exponential distribution. As a direct consequence of this heavy-tailed distribution, the average desorption rate depends upon the time scale of the experiment and the protein surface concentration does not reach equilibrium. Further analysis reveals that the observed anomalous desorption emerges due to the reversible formation of a small fraction of soluble protein multimers (small oligomers), such that each one desorbs from the surface with a different rate. The overall kinetics can be described by a series of elementary reactions, yielding simple scaling relations that predict experimental observations. This work reveals a mechanistic origin for anomalous desorption kinetics that can be employed to interpret observations where low-protein fouling surfaces eventually foul when in long-term contact with protein solutions. The work also provides new insights that can be used to define design principles for non-fouling surfaces and to predict their performance.

\end{abstract}

\maketitle




\section{Introduction}
The interaction of proteins in solution with solid surfaces is a fundamental phenomenon of great importance 
in multiple scientific and engineering disciplines \cite{Ramsden,gray2004,rabe2011,adamczyk2012,wei2014}. 
From a life sciences perspective, adsorption and desorption of proteins at surfaces are key 
players in e.g. organ development, tissue repair, and blood clotting. In biomedical implants and devices, 
controlling or inhibiting irreversible protein adsorption has long been considered an important feature of biocompatible materials and biosensor surfaces \cite{latour2005,fang2005,wei2014,hedayati2019quest}. 
When a biomedical device such as a catheter, or an implant such as a stent or artificial knee comes in contact with body fluids, it is exposed to a myriad of proteins apt to adsorbing into the foreign material and rapidly modifying the surface chemistry \cite{bernard2018biocompatibility}. This adsorbed protein layer can modulate subsequent biological phenomena including blood clotting, bacterial adhesion, and inflammation, which can lead, for example, to failure of blood-contacting medical devices, fouling of contact lenses, and deterioration of biosensor sensitivity. 
Many industrial technologies also rely on controlled protein adsorption for processes related to protein
purification, drug delivery systems, food packaging and storage, and biosensing \cite{Ramsden,firkowska2018}.

Human serum albumin (HSA), the most abundant protein
in blood plasma, and its analogue bovine serum 
albumin (BSA) have been studied extensively as model systems for protein 
adsorption at solid-liquid interfaces. In blood, albumin is responsible for maintaining osmotic pressure \cite{hankins2006role}. 
In the female cervical fluid, it represents $17\%$ of
the total protein \cite{tjokronegoro1975} and it is necessary for sperm to acquire the ability to fertilize an egg \cite{visconti2011ion}.
Several classes of drugs also depend on albumin for binding and transport, including 
antibiotics, anticoagulants, and anti-inflammatory drugs, and albumin is emerging as 
a drug carrier in the treatment of diabetes and cancer \cite{kratz2014}. Further, albumin is used as 
a blocking agent in immunoassays and for coating medical devices to suppress the adhesion of other proteins and bacteria \cite{brokke1991}. The adsorption of albumin on solid interfaces has been studied for multiple substrates including mica\cite{adamczyk2018}, silica \cite{malmsten1994,wasilewska2019}, functionalized gold nanoparticles \cite{dominguez2016}, and model surfaces of varying hydrophobicity \cite{wertz2001}. These studies have typically focused on mass transfer, maximum protein coverage, and the reversibility of protein adsorption. The activation energies for desorption and diffusion have also been investigated in both hydrophobic and hydrophilic surfaces \cite{langdon2012}.  Mathematical models of different complexities have been developed to explain experimental observations \cite{rabe2011,kim2018,adamczyk2018}.

A key feature of irreversible adsorption onto solid surfaces involves surface-induced protein fouling. Thus, a great deal of effort 
has been placed in the design and characterization of advanced materials to impart protein resistance \cite{wei2014,selim2017,hedayati2018,chen2019}. 
Besides their critical applications in biomaterials,
low-fouling surfaces are routinely used in single-molecule biophysics research \cite{roy2008}.  The 
most widely used and best characterized strategy to impart protein resistance to a surface  
consists of coating it with a polyethylene glycol (PEG) brush \cite{banerjee2011}. Experimental evidence shows that the mechanisms of protein interactions with PEG surfaces are highly complex with non-trivial dependence on grafting density \cite{faulon2016}. While  
many studies focus specifically on the rate of surface adsorption, the study of protein desorption  
from low-fouling surfaces is still lacking detailed understanding.

Adsorption and desorption processes are most often quantified in 
terms of the respective kinetic coefficients $\ka$ and $\kd$ 
\cite{fang2005}. This analysis enables experiments to be interpreted using a basic kinetic equation 
\begin{equation}\label{kinetics}
\frac{d\rho}{dt} = \ka-\rho \kd, 
\end{equation}
where $\rho$ is the adsorbed protein surface density and the adsorption 
kinetic coefficient $\ka$ is proportional to the bulk 
protein concentration. This simple kinetic model is expected to predict valid results for 
low surface occupancy so that blocking effects 
can be neglected and both the density of available surface sites 
and the protein solution concentration do  
not change substantially during observation times. \Eq{kinetics} is often sufficient to interpret short-term protein adsorption 
and predicts that surface concentration will converge exponentially to a constant value ($\ka/\kd$) with a characteristic time constant $1/\kd$. 
However,  it often fails to predict long-term (more than $30$ min) surface kinetics. The failure of the kinetic equation 
(\Eq{kinetics}) is sometimes ascribed to a deterioration of the non-fouling behavior due to various effects such as oxidation of the surface 
or the Vroman effect for complex protein mixtures \cite{noh2007}.

In this article, we study the kinetics of BSA on 
a PEG brush-coated surface. The kinetics are characterized in detail by employing single-molecule detection. 
Desorption from the surface is observed to exhibit anomalous behavior that is manifested as a power-law 
distribution in the surface dwell times. This behavior can be accurately explained 
by considering that there is a finite probability for the molecules in solution to reversibly self-associate. 
A simple model considering an equilibrium concentration of multimers in solution and a desorption rate that depends on the number of monomers in the adsorbed particle is proposed to explain 
our results. The predictions from this model are solved semi-analytically and are validated using 
two different surfaces.



%
\section{Materials and Methods}
\subsection{Materials}
PEG silane ($2$-[methoxy(polyethyleneoxy)$6$-$9$propyl] trimethoxysilane), MW $459-591$~Da, 
was purchased from Gelest (Morrisville, PA). 
$\beta$-Mercaptoethanol, catalase from bovine liver, and glucose oxidase were
purchased from Sigma Aldrich (St. Louis, MO). Anhydrous toluene was purchased from MilliporeSigma (Burlington, MA). BSA conjugated to Alexa Fluor $647$ (degree of labeling: $3-6$) and ethanol ($200$ proof $99.5$+\%) were 
purchased from Thermo Fisher Scientific (Waltham, MA). 
$18.2$ M$\Omega$ cm water from a Millipore water purification unit was used for
making all aqueous solutions.


\subsection{Preparation of PEG brush surfaces }
Surfaces functionalized with PEG brushes were constructed via a grafting-to approach \cite{ionov2009,faulon2016}.
Prior to functionalization, fused silica wafers were thoroughly washed with acetone, ethanol, and deionized water and dried with 
ultrapure N$_2$. Wafers were then exposed to oxygen plasma (Plasma Etch, Carson City, NV) for $10$ min. The substrates were subsequently 
incubated in $1$\% v/v PEG silane dissolved in anhydrous toluene. 
The reaction was performed at room temperature for $20$ min to construct PEG brush surfaces. Finally, surfaces were rinsed 
multiple times with toluene and deionized water and dried with ultrapure N$_2$. The higher grafting density surfaces were made 
by incubating the PEG solution for $1$ hour instead of $20$ min.

\subsection{PEG characterization}
The thickness of the dry brush was measured by ellipsometry. For this purpose PEG 
brushes were constructed on Si wafers using the same protocol described above. 
$\langle 100 \rangle$ p-doped $10$-$20$ $\Omega$-cm Si wafers were purchased from  
MSE Supplies (Tucson, AZ). 
Ellipsometry was performed using a J.A. Woollam variable
angle spectroscopic ellipsometer (model VASE-VB-250) and data analysis was done using 
the J. A.Woollam WVASE32 software package. Each surface was spectrally scanned
with an incident angle between $60-80^\circ$, in increments of 5$^\circ$, over 
a wavelength range of $500-900$ nm. The
collected spectra were fit to a three-layer planar model of the solid surface, which accounts for
the refractive index of air ($n = 1.003$), PEG ($n = 1.430$), silicon oxide layer 
($n = 1.457$), and silicon ($n = 3.881$) . 
The dry PEG brush thickness $h$ was obtained and subsequently related to grafting density
$\sigma_{\rm PEG}=\rho_{\rm dry}h N_A/M_w$ where $\rho_{\rm dry}$ is the dry density of 
the PEG monomer repeat unit ($1$ g/cm$^3$), $N_A$ is Avogadro's number, 
and $M_w$ is the average molecular weight of the PEG polymer (500 Da). 
For each surface preparation, dry thickness was measured on three different samples and on two 
different spots in each sample.
The resulting grafting density was $0.15 \pm 0.03$ nm$^{-2}$.
The density of the high grafting density PEG was found to be $0.31 \pm 0.03$ nm$^{-2}$.


\subsection{Imaging buffer}
Imaging was performed in a buffer consisting of 
$50$ mM Tris-HCl (pH $8.0$), $10$ mM NaCl, $0.8$\% glucose, 
$0.15$ mg mL$^{-1}$ glucose oxidase, 34 $\mu$g mL$^{-1}$ 
catalase, and $1$\% $\beta$-mercaptoethanol. This buffer includes 
an enzymatic oxygen scavenging system 
to increase fluorophore stability \cite{roy2008,hedayati2020}.
BSA was added to the imaging buffer to a final concentration of 5 nM.


\subsection{Imaging}
Images were acquired by time-lapse imaging using Nikon NIS-Elements $4.51$ software in an 
objective-type total internal reflection fluorescence (TIRF) custom-built 
microscope equipped with an Olympus PlanApo $100\times$ NA$1.45$ objective and a CRISP 
ASI autofocus system \cite{campagnola2015}. The optical aberrations of the imaging system 
were corrected using a MicAO 3DSR adaptive optics system (Imagine Optic, Orsay,
France) inserted into the emission pathway between the microscope and the
EMCCD camera \cite{gervasi}. Namely, we used the adaptive optics system to correct for 
astigmatism, coma, trefoil, and spherical aberrations. In general, optical elements in the emission path inside the microscope 
introduce different aberrations that compromise the symmetry of the point spread function. The correction of these aberrations 
leads to improved localization precision \cite{clouvel2013}. Fluorophores were excited with a $638$-nm laser 
(DL638-328 050, CrystaLaser, Reno, NV). 
For excitation, an optical density filter with ND=$1.5$ was used and an incident 
angle above the critical angle
was employed. Emission was
collected through the appropriate Semrock bandpass filters 
and the images were acquired in a water-cooled, back-illuminated EMCCD camera 
(iXon DU-$888$, Andor, Belfast, UK) liquid-cooled to $-70^\circ$C with an electronic gain of $60$.  
In order to avoid photobleaching and allow for longer imaging, time-lapsed imaging was 
used whereby an image was obtained every $2$ s over a
total period of $3000$ frames. Exposure time in each frame was limited to $90$ ms using an Uniblitz high-speed optical shutter synchronized with the camera acquisition.

\subsection{Single-molecule detection}
Detection and tracking of individual molecules were performed in MATLAB using the u-track algorithm \cite{jaqaman2008}. The localization precision is governed by the number of detected photons. In our data, the estimated mean localization precision is found to be 10 nm.

\subsection{Measurements of Stokes hydrodynamic radius} 
BSA size was measured by dynamic light scattering (DLS) using a Zetasizer Nano ZS (Malvern) with a $633$ nm laser line. 
Samples were prepared and measured after the solution equilibrated for $20$ min, and the 
same solution was again measured after $90$ min. The time-dependent fluctuations of the  back-scattered light was measured at a fixed angle 
of $175^{\circ}$ in an avalanche photodiode. Samples were maintained at $25 ^{\circ}$C in imaging buffer solution. Acquisition time was set to $70$ s. The intensity autocorrelation functions were analyzed using the Zetasizer Software v.7.13 (Malvern). 
The same software was used to estimate the viscosity for $50$ mM Tris HCl, $10$ mM NaCl aqueous solution at $25 ^{\circ}$C: $\eta=0.894$ mPa s.
The distribution of diffusion coefficients $D$ was directly obtained from the intensity autocorrelation function, which was transformed into a distribution of hydrodynamic radii $R$ using the Stokes-Einstein relation,
\begin{equation}
    R=\frac{k_B T}{6\pi\eta D},
\end{equation}
where $k_B T$ is thermal energy and $\eta$ is the solution viscosity. 
Three independent runs were collected for each sample and the 
obtained distributions of radii were then averaged.

Zeta potential of BSA at pH $8.0$ was also obtained via laser Doppler 
velocimetry in the same Zetasizer Nano ZS instrument. Here, a voltage of $50$ V was applied between electrodes at the edge of the cell separated by $6.1$ mm and the protein electrophoretic mobility $\mu$ is measured. The zeta potential was obtained from Henry's equation \cite{jachimska2008}. We found the zeta potential of BSA to be $-13$ mV.

\section{Results and discussion}
To accurately quantify protein adsorption and desorption on 
PEG-coated surfaces, we imaged individual fluorescently labeled
BSA by total internal 
reflection fluorescence (TIRF) microscopy. 
Proteins were diluted to low
concentrations ($5$ nM) in imaging buffer  
to enable single-molecule detection 
as they adsorb to and desorb from the solid-liquid
interface. The protein solution 
was allowed to equilibrate for at least 1 hour and then
the solution was injected into a 
chamber that has been constructed 
with one surface consisting of a coverslip functionalized with PEG brushes. 
The chamber does not let water evaporate, so that protein 
concentration in solution is constant for the duration of the experiment.
Immediately following injection of the solution, the microscope was focused on the surface, and then time-lapse microscopy videos were collected for 100 min (3000 frames).
Even though the imaging time is long, the concentration of labeled proteins 
in solution is not expected to change substantially because imaging is done in TIRF and, as a consequence, 
photobleaching only affects proteins on the surface. Further, care is taken to 
reduce photobleaching by means of time-lapse imaging and using an oxygen scavenging buffer \cite{hedayati2020}.

\begin{figure}[th] 
\centering
    \includegraphics[width=8.5cm]{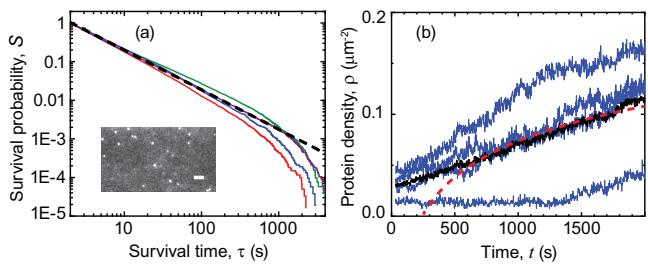}
    \caption{Anomalous surface kinetics. 
    (a) The survival probability of BSA molecules on a PEG surface does not exhibit an 
     exponential decay. Instead it appears to decay as a power law. Four different surfaces that were prepared and measured
under the same conditions, are shown (solid lines) together with a power law function ($\tau^{-\alpha}$, dashed line).
The numbers of molecules detected for at least 2 frames in each surface are $N=28,125$, $N=25,338$, $N=26,757$, and $N=31,579$, totaling overall $111,799$ adsorbed molecules.
    Inset: fluorescent image of adsorbed BSA. Scale bar is 3~$\mu$m.
    (b) The protein surface density increases over time. The $x$-axis represents 
    the time since the protein solution was introduced into the chamber.
    Four experimental surfaces are shown (thin blue lines), together with the 
    density average (black line) and the fit to \Eq{rho} as a dashed red line.
    The initial phase of protein adsorption, accounting for the first 20 s, is not recorded because the microscope is focused during this time taking advantage of adsorbed molecules.
}
  \label{fig01}
\end{figure}

The inset in \fig{fig01}(a)
shows a representative adsorbed BSA image. Individual fluorophores 
are clearly visible above the fluorescent background, making it easy to
detect the exact moments of adsorption and desorption from the surface. 
Adsorbed BSA molecules are observed to be immobile on the PEG surface. 
Supplemental Fig. S1\dag\  shows the distribution of displacements within the trajectories of the 
molecules on the surface both for lag times of a single frame and $5$ frames. The distributions of displacements in $1$ and $5$ frames are indistinguishable and they are governed solely by the localization error. Several groups have previously shown 
that a highly effective mode of surface mobility can be induced by a process known as bulk-mediated diffusion 
\cite{bychuk1994,chechkin2012,skaug2013,rojo2013,weltz2015,campagnola2015,berezhkovskii2015,krapf2016strange}. In this case, 
molecules hop between binding sites by desorbing from one site, diffusing in the bulk, and readsorbing to 
a different site \cite{knight2009,yu2013,skaug2014}. This effect definitely takes place for BSA dynamics, but we do not probe it for two reasons: 
First, the density of labeled proteins on the surface is too high to accurately distinguish between 
a hopping event and the adsorption of a different molecule. Second, we are primarily interested in 
the desorption kinetics, thus we do not investigate bulk-mediated diffusion here.  

Our single-molecule assay allows an evaluation of the kinetic model 
because both $\ka$ and $\kd$ can be directly measured. In particular, \Eq{kinetics} 
predicts the average dwell time for an 
adsorbed particle on the surface to be $\langle\tau\rangle = 1/ \kd$ 
and the dwell times themselves to be random variables drawn 
from an exponential distribution $\psi(\tau)= \kd \exp(-\kd\tau)$.    
\fig{fig01}(a) shows the survival probabilities, $S(\tau)$ of BSA molecules on four 
independent PEG surfaces at the same conditions,
\begin{equation}\label{survival}
S(\tau)=\int_\tau^\infty \psi(t) dt. 
\end{equation}
Contrary to the expectations from the simple kinetic model (\Eq{kinetics}),
the survival probability $S(\tau)$ does not decay exponentially. Counterintuitively, 
it decays as a power law up to a timescale of the order of $1000$~s. 
Namely, the survival probability is observed to scale as $S(\tau)\sim S_0\tau^{-\alpha}$ with 
$\alpha=0.95$. An interesting outcome
of distributions with a power-law tail is that they lack a
characteristic time. In contrast, the apparent desorption rate depends on the time that has lapsed since the system was prepared \cite{schulz2014,krapf2019strange} (in our case, since the 
solution came in contact with the surface). In systems exhibiting heavy-tail distributions, the mean dwell time diverges and a high degree of complexity is expected \cite{stefani2009,jeon2011,weron2017}. 

After $1000$~s, the survival probability decays rapidly  
due to photobleaching. The characteristic photobleacing time depends on the total time 
the protein was exposed to laser excitation. Supplemental Fig. S2\dag\ 
shows the survival probability obtained from imaging without 
time-lapse video (i.e., fluorophores are continuously exposed to laser excitation) 
and using a faster frame rate of $9$ frames/sec. Note that in \fig{fig01}(a), fluorophores 
were only excited during $90$ ms every 2-s frame. Under continuous illumination
the photobleaching decay is observed to occur much earlier. 
The photobleaching decay function is discussed in the Supplementary Information\dag\.  
By probing shorter times
we also find faster kinetics with the same power-law tail
distribution.

The observed anomalous desorption kinetics 
implies that \Eq{kinetics} does not hold and the protein surface density
should not reach a steady state within this long timescale, but it should instead 
increase with time. The surface dwell time (i.e., the desorption time) is given 
by \Eq{survival} as $\psi(\tau)\sim\alpha S_0 \tau^{-(1+\alpha)}$. We can 
write a recurrence relation for the occupation probability $P_{\rm on}$ of an 
individual surface site,
\begin{equation}\label{recurrence}
\frac{dP_{\rm on}(t)}{dt}=A_1 \ka P_{\rm off}(t) - \int_0^t A_1 \ka P_{\rm off}(t') \psi(t-t') dt', 
\end{equation}
where $P_{\rm off}(t)$ is the probability of the site being empty at time $t$, with 
$P_{\rm off}(t)+P_{\rm on}(t)=1$, and $A_1$ is the area of a single site. 
The first term on the right part of \Eq{recurrence} 
has the same meaning as the adsorption in \Eq{kinetics} and the second term  
accounts for a particle being adsorbed at an earlier time $t'<t$ and desorbing at time $t$. 
This recurrence relation is more rigorous than \Eq{kinetics} because it 
relaxes two approximations. The first one is that blocking effects are not necessarily neglected 
and an adsorbed protein can block the adsorption of a new protein. Nevertheless, if the adsorption sites 
are small enough, blocking effects are still negligible. The second and more important one
for this work is that the dwell time of a protein on the surface is not necessarily exponential, i.e., 
the process is not assumed to be Markovian. The physical meaning of the area of an individual site in 
this model is not trivial {\em a priori}. In broad terms, a single site consists of the space available for binding one protein. 
This space can be either the inverse density of surface defects (if proteins are expected to bind onto defects), 
or the area covered by a single particle (if proteins can equally adsorbed anywhere on the surface), 
which depends only on the nature of the surface-particle interactions.

It is possible to solve for $P_{\rm on}(t)$ by use of Laplace transform, which yields 
$P_{\rm on}(t)\sim1-t^{\alpha-1}/c_1$  (see supplementary information\dag\  for mathematical derivation). Note 
that, assuming again low surface occupation, the surface density is proportional to 
the probability of occupation of a single site and thus
\begin{equation}\label{rho}
\rho(t)\sim \frac{1}{A_1} \left(1-\frac{ t^{\alpha-1} }{c_1}\right), 
\end{equation}
where $c_1=A_1 \ka S_0 \Gamma(1-\alpha)\Gamma(\alpha)$. 
 \fig{fig01}(b) shows the density of surface proteins as a function of time for 2000 s in four replicate experiments. The protein density increases with time without reaching a steady state.

\begin{figure}[ht] 
\centering
    \includegraphics[width=5cm]{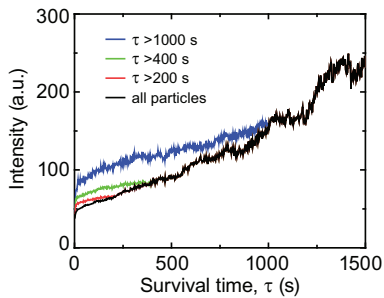}
    \caption{Average intensities of the fluorescent particles as a function
    of the time that lapsed since particle adsorption on the surface. The 
    colored lines show the average intensity for particles that dwell 
    on the surface for longer than a given threshold ($200$, $400$, or 
    $1000$~s in red, green, and blue, respectively).   
}
  \label{fig02}
\end{figure}

A power-law tail in the dwell times can emerge from two different scenarios \cite{krapf2015}: 
(1) A non-stationary process where after capture, the trap strength becomes stronger 
with time and the particle probability of escape decreases \cite{weigel2013} and 
(2) a heterogeneous process involving traps of varying depths 
\cite{scher1975,sokolov2009,burov2011,krapf2013,krapf2019strange}. We are able 
to untangle these effects by measuring the fluorescence intensity of adsorbed particles as a function of 
the time that lapsed since protein adsorption. 
We find that a heterogeneous process
provides the correct interpretation.
The key idea behind this measurement lies in the 
fact that a protein aggregate will be bound more strongly to the surface
than a single monomer while at the same time 
it will exhibit brighter fluorescence emission. 
We do not find evidence for an heterogeneous surface in our data, but intrinsic surface heterogeneities are possible and they would further increase the complexity of the interfacial interactions \cite{langdon2015}. Note that while different protein conformations also lead to heterogeneous binding properties, they are not expected to give rise to different fluorescence intensities.
\fig{fig02} shows, as a black line, the average 
fluorescence intensity for all particles as a function of the time they spent on the surface. 
The average intensity increases with time since adsorption. 
The figure also shows in red, green, and blue 
the average intensities for the particle that survived on the surface at least $200$, 
$400$, and $1000$~s, respectively. Interestingly, the particles that survived longer times on the surface were 
brighter since the time of adsorption. This result suggests that molecules form aggregates in 
solution and then they bind to the surface. Note that, e.g. for particles that survive longer than $1000$ s, 
an increase is also seen up to 1000 s, suggesting that surface-mediated cluster growth can also take place.
Previous DLS evidence has also shown that BSA
suspensions (as well as HSA) contain a
significant fraction of protein dimers \cite{jachimska2008}. The diffusion of BSA at oil-water interfaces has also revealed the presence of BSA dimers and trimers \cite{walder2010}.

\begin{figure}[ht] 
    \includegraphics[width=8.5cm]{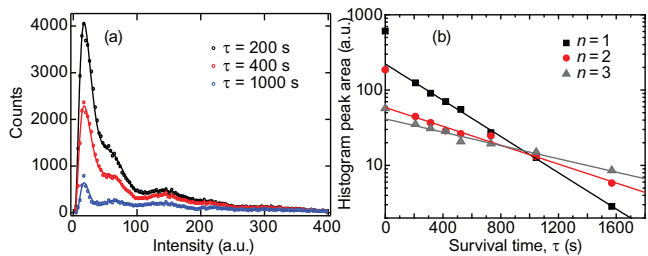}
    \caption{Particle fluorescence intensities show discrete peaks. (a) Histograms of detected particle intensities 
    for particles that survive on the surface longer than $\tau=200$, $400$, and $1000$~s. Several 
    peaks are observed in the histograms. (b) Survival of the first three peaks as a function of time, which 
    correspond to single monomer, dimer, and trimer. 
    The survival is shown as peak area for particles that dwell on the surface longer than $\tau$, instead of actual probability, 
    to preserve the area ratios between different peaks.     
}
  \label{fig03}
\end{figure}

The histogram of particle intensities (\fig{fig03}(a)) displays several well-differentiated 
peaks, in agreement with our hypothesis that particles with different numbers of 
BSA proteins are found on the surface. We observe an intensity peak for single proteins ($n=1$) 
and peaks for dimers  ($n=2$), trimers  ($n=3$), etc. 
We speculate
the first peak (lowest intensity) corresponds to monomers,
the second to dimers, etc.
Following the same methodology 
we employed above, we measure the intensity histograms of particles that dwell on the surface 
for times longer than $\tau$. Three different histograms are shown as examples in \fig{fig03}(a), 
corresponding to $\tau=200$, $400$, and $1000$~s. While the overall survival probability 
of adsorbed particles decays as a power law, the survival probability of each of 
the different peaks decay exponentially, albeit with different characteristic times (\fig{fig03}(b)).

Because particles with longer dwell times are brighter and fluorescent particles on the 
surface have well-defined discrete intensities, we postulate a simple model where proteins in solution 
can aggregate into clusters. Classical biochemistry usually considers that proteins unfold when 
they aggregate. However, unfolding is irreversible for practical purposes and particles 
would sediment out of the liquid phase. Using dynamic light scattering, we verify 
that the measured protein hydrodynamic radius is in steady state. Supplemental Fig. S3a\dag\  shows 
the measured distribution of hydrodynamic radii for a freshly prepared BSA solution and for a solution 
allowed to settle for a period of $90$ minutes. Both distributions are indistinguishable, which provides evidence that the 
solution is in equilibrium and irreversible aggregation does not take place in the bulk phase over the timescale of our experiments. Given that the 
solution is in equilibrium, we assume 
particles can reversibly self-associate into soluble multimers \cite{cromwell2006,Erickson}, and that the distribution of 
the number of proteins ($n$ monomers) in a multimer, $p(n)$, follows a Boltzmann distribution. The addition 
of an extra protein to a cluster requires a free energy $\Delta F$. Thus, $p(n)=[\exp(1/n_0)-1] \exp(-n/n_0)$ with $n_0$ being the 
characteristic number of molecules in a particle (a cluster) in solution \dag. 

Once the particle is on the surface, each constituent 
monomer has the same probability of association to and dissociation from the surface. 
Thus, at a given time, any number of monomers between $0$ and $n$ can be associated to the 
surface,  
\begin{equation}\label{multimer}
\ce{$M_n$ <-->[$na$][$b$] $M_{n-1}$ <-->[$(n-1)a$][$2b$]}
\cdots \ce{<-->[$3a$][$(n-2)b$] $M_2$ <-->[$2a$][$(n-1)b$]
$M_1$ ->[$a$]} M_0
\end{equation}
where $a$ and $b$ are the monomer dissociation and association rates, $M_i$ is a state with $i$ monomers bound to the surface 
and $M_0$ represents a protein that has completely dissociated from the surface. 
Under the approximation $a\ll b$, reaction (\ref{multimer}) has the asymptotic 
long time solution for the probability of being in state $M_0$,
given $n$ monomers in a particle,
\begin{equation}\label{p0}
p_0(\tau|n)\sim 1-\exp(-k_n \tau),
\end{equation}
where $k_n$ is an effective desorption rate coefficient given by
\begin{equation}\label{kn}
k_n=n \frac{a^n}{b^{n-1}}
\end{equation}
for any initial condition between the states $M_1$ and $M_n$ 
(see supplementary information\dag\  for mathematical derivation). 

\begin{figure}[ht] 
    \includegraphics[width=8.5cm]{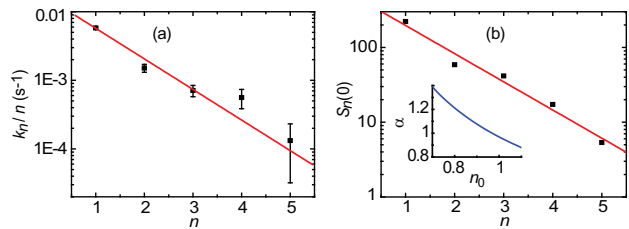}
    \caption{Desorption characterization of particles with different number of monomers. 
    (a) The desorption rate for particles with $n$ monomers is predicted to be $k_n=nb(a/b)^n$. 
    Thus the ratio $k_n/n$ is shown as a function of $n$ to corroborate this prediction. 
    (b) Peak area under each peak extrapolated to $\tau=0$. This value represents 
     the relative amount of adsorbed particles found for each number of monomers. 
     Inset: anomalous exponent $\alpha$ according to \Eq{alpha} when $a/b$ is set to $0.34$.
}
  \label{fig04}
\end{figure}

The survival probability $S_n(\tau)$ for a multimer of $n$ monomers describes the probability
that the particle has not yet reached state $M_0$. Thus, $S_n(\tau) \sim 1-p_0(\tau|n)=\exp(-k_n \tau)$. 
This predicted behavior is in excellent agreement with the exponential decays of the different intensity peaks 
(\fig{fig03}(b)).  Further, the rate $k_n$ is observed to obey the predicted behavior as in \Eq{kn} (\fig{fig04}(a)). 
This measurement yields $a/b=0.34$ and $b=0.008 \, \isec$. Besides 
allowing the computation of $k_n$, an extrapolation of the intensity peaks to $\tau=0$ yields the fraction of molecules in each state, 
i.e., the fraction of single monomers $p(n=1)$, dimers $p(n=2)$, trimers $p(n=3)$, etc., 
that bind to the surface.  \fig{fig04}b shows that these fractions are in good agreement with our assumption of a 
Boltzmann distribution for the number of monomers within a particle in solution, where 
the characteristic number of monomers is found from these data to be $n_0=0.97$.

How does a power-law tail in the surface dwell time distribution emerge? Even though the dwell times of 
particle of a given number of monomers are exponentially distributed, the asymptotic dwell times of a particle of 
unknown $n$ can have a power-law form. Namely, the survival probability is 
\begin{equation}\label{St}
\begin{split}
S(\tau) & =\sum_{n=1}^\infty S_n(\tau)p(n) \\
 & \sim \sum_{n=1}^\infty \exp\left[ -nb\left(\frac{a}{b}\right)^n \tau - \frac{n}{n_0} \right].
\end{split}
\end{equation} 
We analyzed this function numerically and found that, at long times, it converges to a power-law tail 
(Supplemental Fig. S4\dag) for a wide range of $a$, $b$, and $n_0$ parameters. A systematic 
numerical analysis of \Eq{St} reveals $S(\tau)\sim \tau^{-\alpha}$ with 
\begin{equation}\label{alpha}
\alpha=\frac{\gamma}{n_0} \, f\!\left(\frac{a}{b}\right),
\end{equation}
where $\gamma$ is a constant and $f(x)$ is a monotonically increasing function (Supplemental Fig. S5\dag). 
Note that $\alpha$ only depends on the dissociation/association rates ($a$ and $b$) via their ratio. 
By setting $a/b=0.34$ as found from \fig{fig04}(a), we obtain the dependence of $\alpha$ on 
the characteristic number of monomers in a particle, $n_0$ (inset of \fig{fig04}(b)). An anomalous exponent 
$\alpha=0.95$ (as obtained for the tail of the survival probability) is found for $n_0=1.0$. This value is 
in surprisingly good agreement with the value from the relative magnitude of the intensity peaks (\fig{fig04}(b)).      

The observed anomalous surface dynamics is well described by an heterogeneous protein solution with the
heterogeneity being rooted in the proteins forming aggregates in thermodynamic equilibrium. This model
predicts that the distribution of particle sizes in solution, when
measured from dynamic light scattering (DLS) or sedimentation velocity
\cite{harding1995}, should have the same signature of multimers following a Boltzmann distribution. Supplemental Fig.
S3b\dag\  shows the polydispersity of the sample as obtained from DLS measurements. The distribution of hydrodynamic radii exhibits a long tail as expected from an heterogeneous
population. By employing a Boltzmann distribution for
particle sizes,  $p(n)=[\exp(1/n_0)-1] \exp(-n/n_0)$, as before, the
distribution of hydrodynamic radii can be decomposed into
a series of Gaussian peaks. Namely, we 
approximate aggregates of $n$ monomers as a sphere of volume $V_n=nV_1$, where $V_1$ is the volume of a monomer.
Thus, the radius of a cluster is $R_n=n^{1/3}R_1$. From the Boltzmann distribution, the concentration ratio between a multimer of size $n+1$ and that of a multimer of size $n$ is
\begin{equation}\label{ratio}
s=\frac{c_{n+1}}{c_n}=\exp(-1/n_0).
\end{equation}
Therefore, the probability density function (PDF) of radii can be modeled by a series of Gaussian functions with equal width $w$ given by the instrument accuracy,
\begin{equation}\label{rPDF}
p(R)=\sum_{n=1}^{\infty}s^{n-1}A_1 \exp\left[\frac{\left(R-n^{1/3}R_1\right)^2}{2w^2}\right],
\end{equation}
where $A_1$ and $R_1$ are the peak magnitude and radius of the first peak, i.e., those of the monomer. In agreement with
the predictions of our model, a least square fitting of the
radius distribution yields a ratio $s=0.4$, or a characteristic number of molecules of molecules in a cluster, $n_0 = 1.1$.

\begin{figure}[ht] 
    \includegraphics[width=8.5cm]{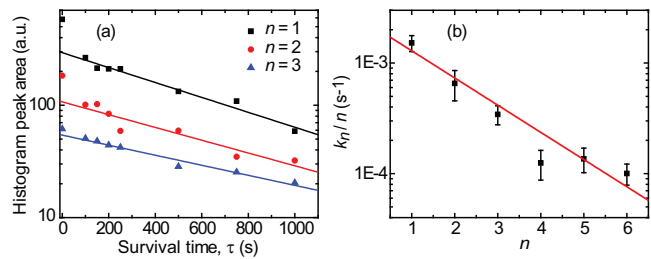}
    \caption{Desorption characterization of particles from a high-density PEG surface. 
    (a) Survival of the first three peaks, i.e. peak area for particles that dwell on the surface longer than a time $\tau$ as a function of the time $\tau$. 
    (b) Desorption rate from high density PEG for particles with $n$ monomers. The ratio $k_n/n$ is shown as a function of $n$ in the same 
    manner as shown in \fig{fig04}(a).
}
  \label{fig05}
\end{figure}

Only two parameters are responsible for the 
anomalous kinetics behavior: the tendency of proteins in solution to oligomerize, described by the characteristic 
number of monomers in a single particle, and the ratio between adsorption and desorption rates of a single monomer 
within a particle on the surface. In order to evaluate our findings on a different set of parameters, we 
modified the PEG conditions to yield a different adsorption/desorption ratio. Specifically, we prepared a new PEG brush 
surface with a higher grafting density.  While the original PEG surface had an average grafting 
grafting density of $0.15\pm0.03$ chains/nm$^2$, the modified surface had an average density of $0.31\pm0.03$ 
chains/nm$^2$, as measured by ellipsometry. 
The multimer model was also found to be in good agreement with the measured kinetics in this surface. A multimodal 
population was detected in the histogram of particle fluorescent intensity and each peak decayed exponentially with time (\fig{fig05}(a)). 
The rate of release from the surface was also observed to obey the behavior predicted by \Eq{kn} (\fig{fig05}(b)) and 
the survival probability exhibits power-law behavior  (Supplemental Fig. S6\dag\ ). However, in this case the ratio 
between monomer binding and unbinding was different than previously found for the low density PEG surface. 
In this case we obtained $a/b=0.57$, which does not allow us to use the small $a/b$ approximation. Thus, a 
model without this approximation is evaluated in this case (Supplemental Fig. S7\dag).

\section{Conclusions}
In order to analyze the adsorption and desorption kinetics on low-fouling surfaces, we studied the kinetics 
of BSA on PEG-coated silica surfaces. Our experimental data show that the release from the surface is 
not governed by an intuitive constant desorption rate and as a consequence, the distribution of dwell times on the 
surface is not exponential. Conversely, the dwell times are drawn from a power-law distribution,causing the apparent desorption rate to depend on the measurement time. 
This anomalous kinetics is found to be rooted in multimers being reversibly formed in solution, where each multimer has 
a desorption rate that depends on the number of monomers.
Detailed understanding of mechanisms of protein accumulation on low-fouling surfaces is essential for design of the next generation of protein resistant surfaces. The model presented in this work describing anomalous desorption kinetics can explain experimental observations of accumulation of proteins on low-fouling surfaces.
%

\section*{Acknowledgements}
We thank Xinran Xu for technical assistance with imaging. This work was supported by the National Science Foundation (award number 1511830).


\bibliography{adrefs} 
\bibliographystyle{rsc} 

\end{document}